\documentclass[newstyle,twocolumn,proceedings]{rmaa}
\usepackage{rmaacite}

\title{Chemical Abundances of Extragalactic \ion{H}{2} regions}
\author{Manuel Peimbert and Antonio Peimbert
  \affil{Instituto de Astronom\'{\i}a, UNAM} }

\fulladdresses{
\item M. Peimbert and A. Peimbert: Instituto de Astronom\'{\i}a,
  UNAM, Apartado Postal 70-264, M\'exico, D.F., 04510, M\'exico
  (peimbert,antonio@astroscu.unam.mx).}
  
\shortauthor{Peimbert \& Peimbert}
\shorttitle{Extragalactic \ion{H}{2} regions}

\keywords{Galaxies: Galaxies: Abundances --- \ion{H}{2} Regions:
Abundances}

\resumen{La determinaci\'on de las abundancias de los elementos pesados en las
regiones \ion{H}{2} extragal\'acticas se obtiene a partir de l\'{\i}neas
excitadas colisionalmente. En la presencia de variaciones de temperatura las
abundancias que se obtienen son cotas inferiores de los valores reales. Para
obtener los valores reales es necesario determinar la estructura de temperatura
de las nebulosas. Discutimos la importancia de obtener intensidades de
l\'{\i}neas de recombinaci\'on de H, He, C, y O para determinar la
composici\'on qu\'{\i}mica de estos objetos. Proponemos que el m\'etodo de
Pagel para encontrar la abundancia de O/H sea calibrado a partir de l\'{\i}neas
de recombinaci\'on observadas y no a partir de modelos de fotoionizaci\'on o de
abundancias obtenidas a partir de l\'{\i}neas excitadas colisionalmente. }

\abstract{The determination of the heavy element abundances from giant
extragalactic \ion{H}{2} regions is based on collisionally excited lines. We
argue that in the presence of temperature variations the abundances determined
are lower limits to the real heavy element abundances. To determine the real
abundances it is necessary to take into account the temperature variations
present in these nebulae. We discuss the relevance of obtaining accurate line
intensities of recombination lines of H, He, C, and O to determine the chemical
composition of extragalactic \ion{H}{2} regions.  We suggest that Pagel's
method to derive the O/H ratio should be calibrated by using recombination
lines instead of photoionization models or abundances derived from
collisionally excited lines. }

\listofauthors{M.~Peimbert \& A.~Peimbert}
\indexauthor{Peimbert, M.}
\indexauthor{Peimbert, A.}

\begin{document}

\maketitle

\section{Introduction}

The chemical abundances of extragalactic \ion{H}{2} regions are paramount for:
a) the study of the heavy elements enrichment of a given galaxy as a function
of position, and b) the enrichment of the heavy elements in the universe as a
function of time by looking at galaxies at different redshifts.

A powerful tool to determine O/H values is provided by Pagel's method
\cite{pag79} which is based on the [\ion{O}{2}] and [\ion{O}{3}] collisionally
excited lines, these lines are very strong and are easily detected in objects
at different redshifts. This method has to be calibrated by matching the
[\ion{O}{3}]/H$\beta$ and [\ion{O}{2}]/H$\beta$ line intensity ratios with
abundances derived from empirical methods or with photoionization models.
Pagel's method might become paramount for the study of the heavy elements
enrichment of the universe as a function of time.

In this review we discuss briefly some observational evidence in favor of the
presence of temperature variations and how large the $t^2$ values are. It has
been shown that in the presence of temperature variations the collisionally
excited lines of a given element provide a lower limit to the abundance of this
element \cite{pei67}.

We argue that Pagel's method has to be calibrated with recombination lines of
\ion{O}{1} and \ion{O}{2}. We propose that the recombination lines of H, He, C
and O should be observed for a set of extragalactic \ion{H}{2} regions with
different heavy element abundances to establish a proper calibration of Pagel's
method.

Recent reviews on the temperature structure of gaseous nebulae are those by
Peimbert (1995, 2002), \scite{pei01}, \scite{est02z}, Liu (2002a, b),
\scite{sta02}, \scite{tor02}, and \scite{pei02w}.

\section{Temperature Structure}

We can characterize the temperature structure of a gaseous nebula by two
parameters: the average temperature, $T_0$, and the mean square temperature
fluctuation, $t^2$, given by

\begin{equation}
T_0 (N_e, N_i) = \frac{\int T_e({\bf r}) N_e({\bf r}) N_i({\bf r}) dV}
{\int N_e({\bf r}) N_i({\bf r}) dV},
\end{equation}

\noindent and

\begin{equation}
t^2 = \frac{\int (T_e - T_0)^2 N_e N_i dV}{T_0^2 \int N_e N_i dV},
\end{equation}

\noindent respectively, where $N_e$ and $N_i$ are the electron and the ion
densities of the observed emission line and $V$ is the observed volume
\scite{pei67}.

For a nebula where all the O is twice ionized we can derive $T_0$ and $t^2$
from the ratio of the [\ion{O}{3}] $\lambda\lambda$ 4363, 5007 lines,
$T_e(4363/5007)$, and the temperature derived from the ratio of the Paschen
continuum to $I({\rm H}\alpha)$, $T_e({\rm Pac}/{\rm H}\alpha)$, that are given
by

\begin{equation}
T_e(4363/5007) = T_0 \left[ 1 + {\frac{1}{2}}\left({\frac{90800}{T_0}} - 3
\right) t^2\right],
\end{equation}

\noindent and

\begin{equation}
T_e({\rm Pac}/{\rm H}\alpha) = T_0 (1 - 1.67 t^2),
\end{equation}

\noindent respectively. A similar equation can be written for the temperature
derived from the ratio of the Balmer continuum to $I({\rm H}\beta)$.

It is also possible to use the intensity ratio of a collisionally excited line
of an element $p + 1$ times ionized to a recombination line of the same element
$p$ times ionized, this ratio is independent of the element abundance and
depends only on the electron temperature. In this review we will adopt the view
that the \ion{C}{2} $\lambda$4267 and the \ion{O}{2} permitted lines of
multiplet 1, are produced by recombination only and consequently that other
mechanisms like radiative transfer, collisions, and fluorescence do not affect
their intensities.

\section{Observational Support for Temperature Variations}

{From} photoionization models of chemically and density homogeneous nebulae it
has been found that $0.002 \leq t^2 \leq 0.03$, with typical values around
$0.005$ (e.g. \pcite{gru92,kin95,per97}).  These low values of $t^2$ indicate
that to assume $t^2 = 0.00$ is a good approximation to derive abundances. But
the observations of many objects, as we will discuss below, indicate
significantly larger values of $t^2$ implying the presence of additional
physical processes not considered by the photoionization models. In what
follows we will mention some of the observations that imply large $t^2$
values. A review of possible causes for these large temperature variations is
presented elsewhere \cite{tor02}.

\subsection{Extragalactic \ion{{\rm H}}{2} regions}

O/H and C/H ratios have been derived from recombination lines in
NGC~5461, NGC~5471, NGC~604, NGC~2363 \cite{est02y}, 30~Doradus, NGC~346 and
LMC~N~11 \cite{tsa02} and 30~Doradus \cite{per02}. The abundances derived from
recombination lines are typically 2--3 times higher than those derived from
collisionally excited lines, to reconcile these differences it is necessary to
adopt $t^2$ values in the 0.023 to 0.10 range.

\scite{lur99f} have computed photoionization models of NGC~2363 and discuss the
variations of the emission spectra obtained with different input
parameters. They find that low metallicity models ($Z=0.10Z_\odot$, the value
derived with $T_e(4363/5007)$ and $t^2 = 0.00$) do not reproduce the observed
features of the spectrum, and conclude that a value of $Z\simeq0.25 Z_\odot$
is in better agreement with the observational data than the usually adopted
value $Z\simeq0.10Z_\odot$.

\scite{gon94} made a detailed observational study of NGC~2363. These
authors based on the determination of $T_e({\rm Pac}/{\rm H}\alpha)$ and
$T_e(4363/5007)$ obtained that $t^2$ is equal to 0.064 for knot A
and 0.098 for knot B in excellent agreement with the best model
by \scite{lur99}.

In general photoionization models predict $T_e(4363/5007)$ values smaller than
observed \cite{sta99,lur99,rel02,pel02,lur02} indicating the possible
presence of an additional heating source not considered by the models.

\scite{pei00} have used the maximum likelihood method, to derive $\tau(3889)$,
$N_e$(\ion{He}{2}), He$^+$/H$^+$, and $T_e$(\ion{He}{2}) based on nine helium
recombination lines of NGC~346. By comparing the $T_e$(\ion{He}{2}) values
with the observed $T_e(4363/5007)$ they obtain $t^2$ in the 0.02 to 0.03 range.

\subsection{Galactic \ion{{\rm H}}{2} regions}

\scite{est99} have found values of $t^2$ in the 0.024 to 0.044 range for the
Orion nebula M8 and M17. The abundance values derived adopting these $t^2$
values are in agreement with chemical evolution models of the Galaxy while the
abundance values derived under the assumption that $t^2 = 0.00$ are not.

The O/H value of the Orion nebula \cite{est98} is 0.02 dex higher than the
average value derived from two recent solar determinations
\cite{hol01,all01}. The Orion nebula value was derived adopting a $t^2 = 0.024$
and a correction of 0.08 dex to consider the fraction of O trapped in dust
grains. This result is in agreement with predictions from models of galactic
chemical evolution.  Alternatively the adoption of $t^2$ = 0.00 implies a
higher O/H value for the Sun than for the Orion nebula.

The O/H abundances for Orion, M8, and M17 derived under the assumption of $t^2
\neq 0.000$ indicate the presence of moderate O/H gradients, while for the O/H
abundances derived under the assumption of $t^2 = 0.000$ the gradients
disappear. This result supports the contention that temperature fluctuations
exist inside gaseous nebulae.

Further discussion and additional arguments in favor of the presence of
significant temperature variations inside galactic \ion{H}{2} regions is
presented elsewhere \cite{pei02w}

\subsection{Planetary Nebulae}

By combining different temperature determinations it has been possible to
determine $t^2$ values in many planetary nebulae. These values are in the 0.00
to 0.15 range, with typical values around 0.04.  Recent reviews on temperature
variations in planetary nebulae have been presented by
Liu (2002a, b), and \scite{tor02}.

\section{Calibration of Pagel's Method to Derive Oxygen Abundances}

The difficulty of measuring $I(\lambda 4363)$ (or any other direct temperature
indicator) led \scite{pag79} to propose an empirical method based on the ratio
of the nebular oxygen lines to $I({\rm H}\beta)$, $R_{23} \equiv I([{\rm
O~II}]\lambda 3727 + [{\rm O~III}]\lambda\lambda 4959, 5007)/ I({\rm H}\beta)$,
to determine the O/H ratio.

There are three different options to calibrate O/H versus $R_{23}$: a) from
photoionization models where the observed nebular lines are matched with those
predicted by the models, b) from abundances based on the observed nebular lines
and $T_e(4363/5007)$ under the assumption of $t^2$ = 0.00., and c) from O
recombination lines.

\subsection{Photoionization models}

This calibration is based on photoionization models where O/H is an input of
the models. Calibrations based on this option have been presented by
\scite{mcc85f}, \scite{dop86}, and \scite{mcg91}. This calibration depends on
the quality of the models. A good model should include an initial mass
function, an age for the stellar burst or the beginning of the star formation,
a star formation rate, and a gaseous density distribution.

The photoionization models not yet include all the physical processes needed to
reproduce all the ratios observed in real nebulae. For example: even the best
photoionization models, those tailored to fit I~Zw~18, NGC~2363 and NGC~346
predict $T_e(4363/5007)$ values smaller than observed \cite{sta99,lur99,rel02}.
The models typically predict $t^2 \approx 0.005$, values considerably smaller
than those derived from observations.

\subsection{Observations of $R_{23}$ and $T_e(4363/5007)$}

The calibrations based on observations adjust the observed $R_{23}$ values with
the abundances derived from $T_e(4363/5007)$ under the assumption that $t^2 =
0.00$. These calibrations depend strongly on the temperature structure of the
nebulae and underestimate the O/H values by factors of about 2 to 3 because,
as mentioned before, $t^2$ is in the 0.023 to 0.10 range.

\subsection{O recombination lines}

There are significant differences between the calibrations of Pagel's method
based on models \cite{mcc85,dop86,mcg91} and the calibrations based on
observations and $T_e(4363/5007)$ \cite{edm84,tor89f,pil00,cas02}. The
differences in the O/H values are in the $0.2$ dex to $0.4$ dex range and could
be due mainly to the presence of temperature inhomogeneities over the observed
volume \cite{cam88,tor89,mcg91,roy96,lur99}. These differences need to be
sorted out if we want to obtain absolute accuracies in O/H of the order of 0.1
dex or better.

We consider that the option to calibrate Pagel's method based on the O
recombination lines is superior to the other two for the following reasons: a)
it is better than the one based on photoionization models because even the best
available models are not yet able to reproduce all the observed emission line
ratios, b) it is better than the option based on the observationally determined
$T_e(4363/5007)$ because the abundances derived from the nebular lines
and $T_e(4363/5007)$ are very sensitive to the $t^2$ value while the O/H values
derived from recombination lines are independent from it.

\section{Discussion and Conclusions}

To be able to constrain the models for the evolution of galaxies as a function
of redshift it is crucial to have good determinations of their heavy element
abundances, and Pagel's method might be the best tool to determine these
abundances.

Some determinations of the O/H values based on recombination lines are already
available for giant extragalactic \ion{H}{2} regions. They yield values in the
$8.2 < log \,\,{\rm O/H} + 12 < 8.8$ range. In the near future it will be possible
to increase the quality of these determinations and to increase the available
range of O/H values.

We propose to calibrate Pagel's method using O recombination lines.  The O
recombination abundances are from 2 to 3 times higher than those derived from
$R_{23}$ and $T_e(4363/5007)$. GRANTECAN can be used for this calibration. A
spectrograph with a resolution higher than 5000 would be needed for this
project.

\end{document}